\documentclass{elsart}
\usepackage[]{graphicx}
\usepackage[]{amsmath, amsfonts, bm}
\usepackage{times}

\journal{SSC}
\begin{document}
\bibliographystyle{elsart-num-ssc}

\begin{frontmatter}



\title{On the upper limit of thermal conductivity GaN crystals}

\author[danil]{B A Danilchenko\corauthref{cor1}}
\corauth[cor1]{Corresponding author}
\ead{danil@iop.kiev.ua}
\author[danil]{I A Obukhov}
\author[tapasz]{T Paszkiewicz}
\author[tapasz]{S Wolski}
\author[jezowski]{A. Je{\.z}owski}

\address[danil]{Institute of Physics, National Academy of Sciences of Ukraine, Prospect Nauki 46, 252650 Kiev, Ukraine}
\address[tapasz]{Faculty of Mathematic and Applied Physics, Rzesz\'{o}w University of Technology, ul. W. Pola 2, PL-35-959 Rzesz\'{o}w Poland}
\address[jezowski]{Institute of Low Temperature and Structure Reasearch, Polish Academy of Sciences, P.O. Box 1410, 50-422 Wroc{\l}aw, Poland}

\begin{abstract}
The maximal value of thermal conductivity $\kappa_{\rm max}$ 
of the perfect wurzite GaN crystal containing isotopes of natural abundance
is estimated. Our upper limit of   $\kappa~=~4800\:{\rm W/Km}$ at $T_{\rm max} = 32\:{\rm K}$ is smaller than calculated by Liu and Balandin  $\kappa=6000\:{\rm W/Km}$ and higher than obtained by Slack \textit{et~al} $\kappa~=~3750\:{\rm W/Km}$.  The phenomenological dependence $\kappa \propto T^{-1.43}$  obtained by Mion~\textit{et~al} for the temperature interval $300-450\:{\rm K}$ is extended to $200-300\:{\rm K}$. For temperatures higher than $T_{\rm max}$ the best fitting of our experimental data to Callaway's formula is obtained for Grueneisen's constant equal to $\gamma = 1.35$.
\end{abstract}

\end{frontmatter}

\section*{}
In heterostructures GaN/AlGaN in electric field of the order of $100 \: {\rm kV/cm}$, the density of dissipated power reaches $10^6 \: {\rm W/cm^2}$ \cite{danil-apl-85-5421,kuzmik2006,barker-2004}. This means that gallium nitride based devices suffer from heating effects that significantly limit the performance of high electron mobility transistors and laser diodes thereby reducing the device's life and reliability. This means that the accurate knowledge of the gallium nitride thermal conductivity is crucial to understanding thermal effects. For this reason, the investigation of the temperature behavior of thermal conductivity coefficient is an actual and reasonable problem.

Recently four groups measured thermal conductivity of GaN in the wide temperature ranges. Luo \textit{et al} \cite{luo2001} have measured the thermal conductivity of both LEO and HVPE single crystals of GaN over the temperature range $60-300\:{\rm K}$. Slack~\textit{et~al}~\cite{slack2002,morelli2002} performed measurements in the temperature interval $11-300\:{\rm K}$. We measured thermal conductivity of massive high pressure growth dislocation-free crystalline sample containing about $10^{20}$ per $\rm{cm}^{-3}$ point defects in the temperature interval $4.5-300\:{\rm K}$ and performed a preliminary analysis of temperature dependence of it \cite{jezowski2003}. We also measured heat capacity of GaN for temperatures between $5 {\rm K}$ and $300 {\rm K}$, and as a result we established the value of Debye's temperature of GaN \cite{danil2006}. These measurements also allow us to obtain the temperature dependence of the phonon mean free path $L_{\rm eff}$. Measurements of thermal conductivity for temperatures higher than $300\:\rm{K}$ were performed by Mion \textit{et al} \cite{mion2006}. 


Slack \textit{et al} compared their experimental results to the Callaway formula considering separately the contributions of longitudinal and transverse phonons. Therefore, their fitting procedure relays on six fitting parameters, namely two Debye's temperatures 
$\theta_{\rm L}$, $\theta_{\rm T}$ 
Grueneisen's constants 
$\gamma_{\rm L}$, $\gamma_{\rm T}$ 
and constants $b_{\rm L}$, $b_{\rm T}$ related to the Umklapp processes (L and T stands for longitudinal and transverse phonons, respectively). The single set of their free adjustable parameters contained also the relaxation rate for the normal processes 
$\tau _N^{ - 1} $. 
Liu et Balandin \cite{balandin2005} considered thermal conductivity GaN. They argued that in the case of GaN specimens the normal processes do not influence the thermal conductivity because of strong scattering on point defects.  For this reason we use Calaway's formula 
with  $\kappa_2=0$. The results of measurements of thermal conductivity on wurzite GaN samples \cite{luo2001,slack2002,morelli2002,jezowski2003} were analyzed by Kamatagi \textit{et al} \cite{kamatagi2007} employing Holland's and modified Callaway's models. A good representation of the temperature dependence of $\kappa$ was obtained with a single set of phonon-phonon scattering prameters. The purpose of this study is different than Kamatagi \textit{et al} \cite{kamatagi2007}. Having in mind the needs of high-power electronic devices manufactured from GaN we shall establish the \emph{upper} limit of thermal conductivity for $T>T_{max}$. We base our approach on the standard Callaway formula and use only \emph{two} free adjustable parameters. 


In dielectric and semiconducting crystals heat is carried mainly by acoustic phonons. As a rule, the group velocities of such phonons are much greater than those of optical phonons. Therefore, heat flow in semiconducting crystals is limited by various mechanisms of scattering of acoustic phonons. There are two principal kinds of phonon scattering in crystals, namely resistive and normal processes \cite{Berman1976}. They are characterized by the corresponding relaxation times $\tau_{\rm R}$  and $\tau_{\rm N}$. Normal processes do not change the total momentum of phonons. Such processes influence the heat flow only indirectly. The resulting phonons of higher energy are scattered more intensively on point defects and have greater probability of participating in Umklapp processes. In the perfect crystals (dislocation and chemical impurity-free samples), the relevant intrinsic resistive scattering is due to three--phonon Umklapp with relaxation time $\tau_{\rm U}$, the point defect scattering with characteristic time $\tau_{\rm P}$,  and the boundary scattering  $\tau_{\rm B}$. The characteristic time $\tau_{\rm P}$  is determined mostly by the crystal natural isotope composition.  

The total thermal conductivity  $\kappa(T)$ can be written as \cite{Callaway1959}
\begin{equation}
 \label{eq:kappa}
		\kappa = \kappa_1 + \kappa_2,
\end{equation}
where $\kappa _1 (T)$ and $\kappa _2 (T)$  are defined by 
\begin{equation}
\label{eq:kappa1-def} 
 \kappa _1 (T) = \left( {\frac{{k_B }}{\hbar }} \right)^3 \frac{{k_B }}{{2\pi ^2 \bar v}}T^3 \int\limits_0^{\frac{{\theta _D }}{T}} {\frac{{\tau _c x^4 e^x }}{{\left( {e^x  - 1} \right)^2 }}} dx,
\end{equation} 
\begin{equation}
\label{eq:kappa2-def}
\kappa _2 (T) = \left( {\frac{{k_B }}{\hbar }} \right)^3 \frac{{k_B }}{{2\pi ^2 \bar v}}T^3 \left\{ {\int\limits_0^{\frac{{\theta _D }}{T}} {\frac{{\tau _C }}{{\tau _N }}x^4 e^x \left( {e^x  - 1} \right)^{ - 2} } dx} \right\}^2 \left[ {\int\limits_0^{\frac{{\theta _D }}{T}} {\frac{{\tau _C }}{{\tau _N \tau _R }}x^4 e^x \left( {e^x  - 1} \right)^{ - 2} } dx} \right]^{ - 1},
\end{equation}
where ${\rm k_B}$ and $\hbar$  are Boltzmann's and Planck's constants respectively, $\bar{v}$  is the mean velocity of phonons,  $ \theta_D$ is Debye's temperature, $T$ is the ambient temperature,  
$x = \hbar \omega /k_B T$,  
$\omega$ is the phonon frequency,   
$\tau _C^{ - 1}  = \tau _R^{ - 1}  + \tau _N^{ - 1} $, 
and  
$\tau _R^{ - 1}  = \tau _U^{ - 1}  + \tau _P^{ - 1}  + \tau _B^{ - 1} $.  


When 
$\tau _N  >  > \tau _R $,
the combined relaxation time 
$\tau _c  \approx \tau _R $
and the term  $\kappa_2$ (\ref{eq:kappa2-def}) is negligibly small. The generally accepted expression for Umklapp phonon scattering rate is (cf. e.g.~\cite{slack2002})
\begin{equation}
\label{eq:tau_u-def}
\tau _U^{ - 1}  = \frac{{\hbar \gamma ^2 T\omega ^2 }}{{M\bar v^2 \theta _D }}\exp \left( { - \frac{{\theta _D }}{{bT}}} \right),
\end{equation}
where $\gamma$  is Grueneisen's constant, $M$ is the mass of GaN molecule,  and is $b$ is a fitting parameter. As a rule for crystalline solids, 
$2 \le b \le 3$
\cite{Berman1976}.


For phonons scattered by isolated defects of mass different from that of the host in an otherwise perfect crystal, Tamura derived a scattering rate \cite{Tamura1984}
\begin{equation}
\label{eq:tau_P}
\left[ {\tau_P (x)} \right]^{ - 1}  = \frac{{Vk_B^4 \tilde{g }}}{{12\pi \hbar ^4 \bar v^3 }}x^4 T^4,
\end{equation}
where $V$ is the volume of the elementary cell, the parameter $\tilde{g}$ is
\begin{equation}
	\tilde{g}  =\frac{
	\sum\limits_\sigma {g(\sigma)}\bar{M}^{2}(\sigma)
	}
	{
	\left[\sum\limits_{\sigma'}\bar{M}({\sigma'})\right]^2
	}  , 
\label{eq:g-tilde}
\end{equation}
with $g(\sigma)$ is the constant which express the strength of isotope effecs. It depends on mass $M_{i}(\sigma)$ and fraction $f_{i}(\sigma)$ in the 
ith isotope of $\sigma$ atom as 
\begin{equation}
g(\sigma)=\sum\limits_i f_{i}(\sigma)\left[1-\frac{M_{i}(\sigma)}{\bar{M}(\sigma)}\right]^{2}.
\end{equation}
Above $\bar{M}(\sigma)=\sum\limits_i f_{i}(\sigma)M_{i}(\sigma)$ is the average mass of $\sigma$ atom.

The boundary scattering is important for low temperatures and does not influence thermal conductivity in the intermediate and high temperatures. The related scattering rate 
$\tau _B^{ - 1} $ is 
\begin{equation}
\label{eq:tau_b}
	\tau _B^{ - 1}  = \frac{{\bar v}}{{L_{\rm eff} }}
\end{equation}
where $L_{\rm eff}$ is the effective phonon mean free path. 
For 
$T  \ll{\rm \theta_D}$ 
it is of the order of the cross--sectional dimensions. A precise determination of $L_{\rm eff}$ is crucial for correctly describing the $\kappa(T)$  curve in the low temperature range.


As we have recently established, for GaN, the Debye temperature 
$\theta  = 365\,{\rm K}$~\cite{danil2006}, 
which is lower than used by other authors~\cite{slack2002,morelli2002,balandin2005,balandin2002jap,balandin2001apl}. It is therefore obvious that one should revise the fitting procedure. We calculate $\tilde{g}$ (Eq.\ref{eq:g-tilde}) accounting for natural composition of isotopes of Ga and N. Our calculations show, in agreement with ref. \cite{morelli2002}, that the contribution of nitrogen isotopes to $\tilde{g}$  is ten times smaller than the contribution of isotopes of gallium.


Since other defects always present in available specimens diminish the thermal conductivity coefficient, we obtain the upper limit of thermal conductivity of GaN specimens with the natural composition of isotopes. The Grueneisen parameter for hexagonal GaN was calculated by {\L}opuszy{\'n}ski and Majewski \cite{lopuszynski2007}. Their value 
$\gamma  = 1.28$
is bigger that used by other authors~\cite{morelli2002,balandin2005,iwanag2000}. For temperatures higher than $T_{\rm max}$ (corresponding to $\kappa_{\rm max}$), the best fitting is obtained for $\gamma = 1.35$, which is close to the calculated value~\cite{lopuszynski2007}. As a result of fitting the Callaway formula to our experimental results (with $\tau _N  \gg \tau _R $), we found $b = 2.5$.  In Table \ref{tab:table1} we collected the values of the above parameters. 

\begin{figure*}[htbp]
	\centering
		\includegraphics{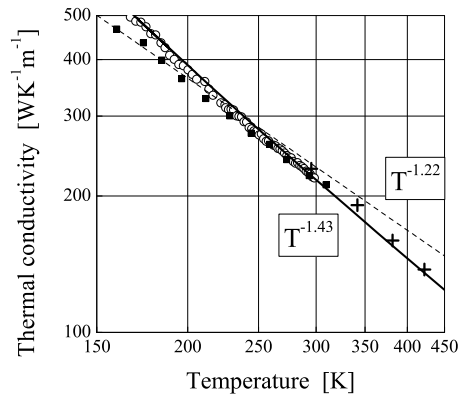}
	\caption{Thermal conductivity  $\kappa(T)$ of GaN for $T>T_{\rm max}$. Open circles -- our results \cite{jezowski2003}, crosses are taken from ref. \cite{mion2006} for sample with the low dislocations density, filled squares -- ref. \cite{slack2002}. Solid line -- the present calculations, dotted line -- $\kappa  \propto T^{ - 1.22}$ ref. \cite{slack2002}. }
	\label{fig:therm_cond1}
\end{figure*}

\begin{table*}[htbp]
	\centering
		\begin{tabular}{|c|c|c|c|c|}
		\hline
			$\bar{v}\:{\left[cm/s\right]}$ & $\gamma$ & $\tilde{g}$ & $b$ & ${\rm \theta_D\left[K\right]}$\\ \hline
			$5.1 \times 10^5$ & $1.35$ & $1.37 \times 10^{-4}$ & $2.5$ & $365$ \\ \hline
		\end{tabular}
	\caption{Single parameter set based on Callaway's model used to represent the temperature dependence of the thermal conductivity $\kappa$}
	\label{tab:table1}
\end{table*}
Extension of temperature dependence calculated by us is in excellent agreement with  $\kappa  \propto T^{- 1.43}$ 
 obtained in ref. \cite{mion2006} in the temperature interval $300 - 450{\rm K}$ for a specimen containing dislocations of the lowest density. This means that the phenomenological expression
$\kappa  = \kappa (300{\rm K}) \times \left( {T/300 {\rm K}} \right)^{ - 1.43} $
proposed in ref. \cite{mion2005dis} can be also used in temperature interval $200-450 {\rm K}$. One should stress that this phenomenological expression differs from the results set forth in ref. \cite{morelli2002}  (cf. Fig. \ref{fig:therm_cond1}). The observed numerical values of $\kappa$  and the temperature slope of  $\kappa(T)$ indicates that $3{\rm \omega}$  dynamic method of thermal conductivity measurements \cite{mion2006} and the axial stationary heat flow method measurements \cite{jezowski2003} yield the same temperature dependence, hence these two experimental techniques should be considered as supplementary. 

\begin{figure*}[htbp]
	\centering
		\includegraphics{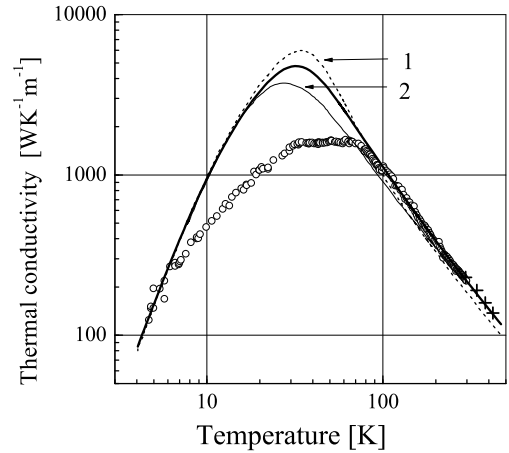}
	\caption{Thermal conductivity of gallium nitride in the temperature interval $5-450 {\rm K}$. Open circles -- results for our best samples \cite{jezowski2003}, thick solid line -- present calculation, crosses -- ref. \cite{mion2006}, curve No. 1 - ref. \cite{balandin2005}, curve No. 2 -- ref. \cite{morelli2002}. }
	\label{fig:therm_cond2}
\end{figure*}

In Fig. \ref{fig:therm_cond2} we present the results of our calculations for 
$3 \le T \le 500\:{\rm K}$~\cite{jezowski2003}
and compare them with the results of calculations presented in refs. \cite{morelli2002,mion2006,balandin2005}. In the low temperature limit, where the boundary scattering is essential, the calculated $\kappa$  was fitted to our experimental results using the value of  $L_{\rm eff} = 0.08\:{\rm cm}$. This value is close to the low temperature mean free path of acoustic phonons reported in ref. \cite{danil2006}.

The mean value of velocity is calculated using the set of elastic constants reported by Polian \textit{et al} \cite{polian1996}. The obtained value of   $\bar{v}$ is close to velocity  $\bar{v}=6\times 10^5\:{\rm cm/s}$ measured in our experiments on propagation of phonon pulses at helium temperatures \cite{danilch1999}. 


Our upper limit of $\kappa$ ($\kappa_{\rm max}=4800\:{\rm  W/Km}$ at $T_{\rm max} = 32\:{\rm K}$) is smaller than calculated by Liu and Balandin  $\kappa_{\rm max}=6000\:{\rm  W/Km}$ \cite{balandin2005} and higher than value $\kappa_{\rm max}=3750\:{\rm  W/Km}$ announced in ref. \cite{morelli2002}.  


In conclusion, for hexagonal GaN crystals containing isotope atoms and otherwise perfect crystal, we obtained an acceptable representation of all our data for temperatures between $3\:{\rm K}$ and $500\:{\rm K}$. Using Callaway's model with  $\kappa_2=0$ ($\tau_{\rm N}\gg \tau_{\rm R}$), we adjusted two free parameters ($\gamma$  and $b$) in the combination of the two scattering mechanisms considered, with the scattering rates for isotopic and boundary scattering given in Eqs. (\ref{eq:tau_P}) -- (\ref{eq:tau_b}). The remaining parameters were calculated or deduced from independent experiments. We established that thermal conductivity for temperatures greater than $T_{\rm max}$ is determined by scattering by point mass defects and the Umklapp processes. In the temperature range $200-500\:{\rm K}$ we justified the empirical dependence $\kappa  \propto T^{ - 1.43} $
observed by Mion \textit{et al} \cite{mion2006}. 

\section*{Acknowledgments}
This work was supported by the Program of Targeted Initiatives established by STCU  USA and NASU Ukraine under Grant No. 3922. 


\begin{thebibliography}{10}
\expandafter\ifx\csname url\endcsname\relax
  \def\url#1{\texttt{#1}}\fi
\expandafter\ifx\csname urlprefix\endcsname\relax\def\urlprefix{URL }\fi

\bibitem{danil-apl-85-5421}
B.~A. Danilchenko, S.~E. Zelensky, E.~Drok, S.~A. Vitusevich, N.~Danylyuk, S.
  V.~Klein, H.~L{\"u}th, A.~E. Belyaev, V.~A. Kochelap, Appl. Phys. Lett. 85
  (2004) 5421.

\bibitem{kuzmik2006}
J.~Kuzm{\'i}k, S.~Bychikhin, D.~Pogany, C.~Gaquiere, E.~Morvan, J. Appl. Phys.
  99 (2006) 123720.

\bibitem{barker-2004}
J.~M. Barker, D.~K. Ferry, S.~M. Goodnick, D.~D. Koleske, A.~Allerman, R.~J.
  Shul, J. Vac. Sci. Technol. 22 (2004) 2045.

\bibitem{luo2001}
C.~Luo, D.~R. Clarke, J.~R. Dryden, J. Electron. Mater. 30 (2001) 138.

\bibitem{slack2002}
G.~A. Slack, L.~J. Schowalter, D.~Morelli, J.~A. Freita~Jr, J.Cryst. Growth.
  246 (2002) 287.

\bibitem{morelli2002}
D.~T. Morelli, J.~P. Heremans, G.~A. Slack, Phys. Rev. B 66 (2002) 195304.

\bibitem{jezowski2003}
A.~Je{\.z}owski, B.~Danilchenko, M.~Bo{\'c}kowski, I.~Grzegory, S.~Krukowski,
  T.~Suski, T.~Paszkiewicz, Solid State Com. 128 (2003) 69.

\bibitem{danil2006}
B.~A. Danilchenko, T.~Paszkiewicz, S.~Wolski, A.~Je{\.z}owski, T.~Plackowski,
  Appl. Phys. Lett. 89 (2006) 061901.

\bibitem{mion2006}
C.~Mion, J.~F. Muth, E.~A. Preble, D.~Hanser, Appl. Phys. Lett. 89 (2006)
  092123.

\bibitem{balandin2005}
W.~Liu, A.~A. Balandin, J. Appl. Phys. 97 (2005) 073710.

\bibitem{kamatagi2007}
M.~Kamatagi, N.~Sankeshwar, B.~Mulimani, Diamond {\&} Related Materials 16
  (2007) 98.

\bibitem{Berman1976}
R.~Berman, Thermal Conductivity in Solid, Oxford University Press, Oxford,
  1976.

\bibitem{Callaway1959}
J.~Callaway, Phys. Rev. 113 (1959) 1046.

\bibitem{Tamura1984}
S.~I. Tamura, Phys. Rev. 30 (1984) 849.

\bibitem{balandin2002jap}
J.~Zou, D.~Kotchetkov, A.~A. Balandin, D.~I. Florescu, F.~H. Pollak, J. Appl.
  Phys. 92 (2002) 2534.

\bibitem{balandin2001apl}
D.~Kotchetkov, J.~Zou, A.~A. Balandin, D.~I. Florescu, F.~H. Pollak, Appl.
  Phys. Lett. 79 (2001) 4316.

\bibitem{lopuszynski2007}
M.~{\L}opuszy{\'n}ski, J.~Majewski, Ab initio calculations of third-order
  elastic constants and related properties for selected semiconductors,
  arXiv:cond-mat/0701410v2 (Apr 2007).

\bibitem{iwanag2000}
H.~Iwanag, A.~Kunishig, S.~Takeuchi, J. Mat. Sc. 35 (2000) 2451.

\bibitem{mion2005dis}
C.~Mion, Investigation of the thermal properties of gallium nitride using the
  three omega technique, Ph.D. thesis, North Carolina State University Raleigh
  (2005).

\bibitem{polian1996}
A.~Polian, M.~Grimsditch, I.~Grzegory, J. Appl. Phys 79 (1996) 3343.

\bibitem{danilch1999}
B.~A. Danilchenko, M.~Bo{\'c}kowski, I.~Grzegory, V.~Guzenko, T.~Paszkiewicz,
  T.~Suski, Physica B 263--264 (1999) 727.

\end{thebibliography}

\end{document}